\documentclass{ws-mpla}

\newcommand{\bea}{\begin{eqnarray}}
\newcommand{\eea}{\end{eqnarray}}
\newcommand{\non}{\nonumber}

\begin{document}

\markboth{Massimiliano Rinaldi}
{Aspects of Quantum Gravity in Cosmology}

%%%%%%%%%%%%%%%%%%%%% Publisher's Area please ignore %%%%%%%%%%%%%%
\catchline{}{}{}{}{}
%%%%%%%%%%%%%%%%%%%%%%%%%%%%%%%%%%%%%%%%%%%%%%%%%%%%%%%%%%%%%%%%%%%

\title{Aspects of Quantum Gravity in Cosmology
}

\author{\footnotesize MASSIMILIANO RINALDI}

\address{naXys, Namur Center for Complex Systems, University of Namur\\
8 Rempart de la Vierge, B5000 Belgium\\
mrinaldi@fundp.ac.be}

\maketitle

%\pub{Received (Day Month Year)}{Revised (Day Month Year)}

\begin{abstract}
We review some aspects of quantum gravity in the context of cosmology. In particular, we focus on models with a phenomenology accessible to current and near-future observations, as the early Universe might be our only chance to peep through the quantum gravity realm.

\end{abstract}

\ccode{PACS Nos.: include PACS Nos.}

\section{Introduction}	

The Einstein field equations are classical, in the sense that they do not contain the Planck constant, and it is a matter of fact that the gravitational interactions have infinite range, and are extremely weak on short distances. At first glance, the question of quantizing gravity appears nonsensical precisely because gravitational forces are negligible at the scale of the quantum world. This is certainly true in most part of the Universe, but there are important exceptions that require the extension of general relativity (GR) to some quantum theory of gravitation. Indeed, GR predicts the existence of singularities, namely spacetime regions where curvature and energy density diverge. This situation occurs in two families of solutions to the Einstein equations. The first  describes the collapse of spherical shells of matter that ends with the  formation of a black hole, with a singularity shielded by an event horizon. The second class of solutions models the large scale structure of the Universe and its time evolution. Since the Universe appears to be isotropic, homogeneous and expanding, the Einstein equations predict that a singularity occurred in the past. There is a crucial difference between these two singularities. The one inside of a black hole is always hidden behind an event horizon that prevents from observing what happens around it. In opposition, in the cosmological case, we actually \emph{live inside} the event horizon (although one should be a bit careful in defining horizons in cosmology \cite{Hawking:1973uf,Mukhanov:2005sc}). Therefore, at least in principle, we have observational access to any instant arbitrarily close to the initial singularity. This makes early Universe the unique laboratory with access to the regime where GR is pushed towards its own limits of validity. A simple estimation shows that quantum gravitational effects  become important when the energy density is about $10^{19}$ GeV, that was presumably reached when the Universe was about $10^{-43}$ seconds old. This value is 19 orders of magnitude larger that the energy density reached at LHC. Therefore, the only hope to observe a glimpse of quantum gravitational effects relies upon the study of the early Universe.

In this paper, we review some ideas conceived to tackle the problem of quantum gravity, especially in the context of cosmology. As the subject is extremely vast, we decided to contain ourself only to some aspects. Thus, we have chosen a path that begins from the quantization of fields on a curved background, then goes on towards the transplanckian region, and ends up with some more speculative high-energy models, with the aim of keeping a firm foot on the observational ground. In fact, the dramatic advances in technology of the last few years allow for measurements of cosmological parameters with unprecedented precision, and  great expectations come along with space-based missions currently under operation, like Planck \cite{sitePL}, and under implementation, like Euclid \cite{EditorialTeam:2011mu,site}. For this reason, we think that the phenomenological aspects of any quantum theory of gravitation is of the utmost importance when applied to cosmology.

The plan of this review is the following. After a survey of various lines of research in Sec.\ 2, we focus on the semiclassical theory in Sec.\ 3. In Sec.\ 4 we consider the regime where  the semiclassical theory needs to be modified by planckian effects. We also report on some phenomenological models of quantum gravity, focussing our attention to non-commutative geometry. We conclude in Sec.\ 5 with some remarks and open problems.

\section{Lines of research}

The quantization of the gravitational field is a longstanding problem. The first attempts (by Fierz, Pauli and others) were based on the canonical quantization of the gravitational field fluctuations over a  curved fixed background, inspired by the quantization procedure of electromagnetism.  This approach, often called semiclassical, finds its own limits in the fact that GR is not renormalizable, as  established by Veltman and 't Hooft in the seventies. Despite this, the semiclassical theory has lead to many important discoveries on the side of quantum gravity phenomenology, and laid the foundations for extensions of GR, that ultimately led to string theory together with its  description of the Universe known as string cosmology. The main feature of string cosmology is that the initial singularity is traded for a bounce that happens after a contracting phase, which is related to the expanding phase via a fundamental symmetry of string theory called T-duality \cite{Gasperini:2002bn}. The price to pay is that the details of the bounce depends on non-perturbative aspects of string theory that are still unclear. Also, the observational predictions seem to be not in agreement with the present data. String cosmology has lead to many spin-off theories, most notably brane cosmology.  This model explores the possibility of extra warped spatial dimensions \cite{Brax:2004xh}, modeling our Universe as a membrane sitting in a de Sitter space \cite{Randall:1999ee,Randall:1999vf} or moving in more general ones \cite{Birmingham:2001dq,Kraus:1999it}. The implementation of brane cosmology with  T-duality was also investigated \cite{Corradini:2005pm,Rinaldi:2004hh,Rinaldi:2003is}.

A more ambitious way of quantizing gravity follows the path of canonical quantization and the construction of a Hilbert space that carries the representations of operators associated to the metric tensor, in a background free fashion. This line of research, started by Dirac, Bergman and, later, Arnowitt, Deser, and Misner, lead deWitt and Wheeler to write down a formal Schr\"odinger-like equation for the theory \cite{dewitt,wheeler}. This results was much improved in the following years by loop quantum gravity (LQG), that resolves many ill-defined aspects of the Wheeler-deWitt equation. An important remark is that, in cosmology, the quantization procedure is performed after that homogeneity and isotropy symmetries are imposed. In this sense, loop quantum cosmology (LQC) does not exactly overlap with LQG. One of the main result of LQC is the construction of a cosmological model that does not incur into a singularity \cite{Banerjee:2011qu,Ashtekar:2011ni,Bojowald:2011zz}. The picture that emerges is that of a bouncing cosmology with a super-inflationary phase (i.e. a phase with $\dot H>0$, where $H$ is the Hubble parameter) that follows the bounce. However, this phase lasts for a very short time, so a standard inflationary mechanism is still necessary to account for observations. 

There are several other models that offer alternatives to the direct quantization of gravity. Recently, Ho\v{r}ava has proposed a power-counting renormalizable theory of gravity, based on an anisotropic scaling at high energy \cite{Horava:2009uw}. Essentially, the fundamental hypothesis is that time and space do not scale in the same way, according to the scheme $t\rightarrow b^{z}t$, $x^{i}\rightarrow bx^{i}$, where $z$ is called critical (Lifschitz) exponent and $b$ is an arbitrary constant. By adding higher spatial curvature terms to the standard Einstein-Hilbert action, one can construct a model where, at high energy $z\geq 3$, which makes the theory power-counting renormalizable, while at low energy $z=1$. Local Lorentz invariance is preserved in the infrared (IR), and it is broken in the UV. The original formulation of this model suffered from un unwanted ghost scalar field, that persisted also in the IR \cite{Koyama:2009hc,Blas:2009yd}. To remove this anomalous degree of freedom one needs to add new terms in the action, that are basically formed by combination of a vector field, orthogonal to constant time surfaces, and its derivatives \cite{Blas:2009qj}. In this form, the Ho\v{r}ava-Lifschitz theory becomes very similar to the ``Einstein-aether'' theory proposed by Jacobson many years before as a vector-tensor theory of gravity \cite{Jacobson:2010mx}. Both theories offer non-singular solution to the cosmological equations \cite{Calcagni:2009ar,Brandenberger:2009yt} and the horizon problem is solved without recurring to inflation \cite{Mukohyama:2009gg}. 

An interesting aspect of the Ho\v{r}ava-Lifschitz theory is that the spectral dimension decreases from $d_{s}=4$ in the IR to $d_{s}=2$ in the UV \cite{Horava:2009if}. As remarked by Carlip, this features occurs in many independent models of quantum gravity \cite{Carlip:2009km,Carlip:2009kf}, where both spectral and geometric dimension collapse from four to two in the vicinity of the Planck scale. An explanation of this phenomenon, often dubbed ``asymptotic silence'' is based on the analysis of the Raychaudhuri equation coupled to quantum fluctuations. Surprisingly, it turns out that, at least in two dimensional conformal dilaton gravity, fluctuations of planckian energy tend to focus the light cone at each space-time point into a space-like line  \cite{Carlip:2011tt}, and there is some evidence that this mechanism works also in four-dimensional spacetime. If this were true, this model would open the possibility that the very early Universe was effectively two-dimensional. In this case, the observational signatures might appear in the form of a maximum frequency for primordial gravitational waves \cite{Mureika:2011bv}. In contrast, such a scenario seems to introduce oscillations in the scalar and tensor perturbation spectra that are not practically observable \cite{Rinaldi:2010yp}.

To conclude this section, we wish to mention that some researchers believe that the Einstein field equations need not to be quantized. Rather, they are interpreted as an equation of state analogous to the second law of thermodynamics \cite{Jacobson:1995ab,Padmanabhan:2003gd}. Although very intriguing, this idea has not been developed in the context of cosmology.

\section{Semiclassical theory}

The semiclassical approach to quantum gravity deals with quantum fields defined on a classical background \cite{ParkerToms,BirrellandDavies}. The main idea is to couple the \emph{quantum} energy momentum tensor of the matter fields to gravity via the usual Einstein equation
\bea\label{semiclassical}
R_{\mu\nu}-{1\over 2}g_{\mu\nu}R=8\pi G \langle T_{\mu\nu}\rangle.
\eea
Here, $G$ is the Newton's constant and $\langle T_{\mu\nu}\rangle$ represents the expectation value of the quantum stress tensor of some matter field, defined on a specified quantum state.  As it happens in flat space,  the  expectation value of the stress-tensor is a UV divergent quantity. However, on curved spacetimes, the divergence cannot be simply cured by normal ordering, as new infinities occur, due to the curvature itself. Therefore, one needs to add counter-terms, to the usual Einstein-Hilbert action, that take the form of powers of $R$, $R_{\mu\nu}$, $R_{\mu\nu\rho\sigma}$, and derivatives thereof. These terms then appear on the left-hand side of Eq.\ (\ref{semiclassical}), and the equations of motion become extremely complicated even in the case of highly symmetric backgrounds, such as FLRW. Despite these difficulties, one might hope that the quantum backreaction can avoid the formation of a cosmological singularity. In fact, the renormalized energy density $\rho$ and pressure $p$ might allow for the violation of the energy condition $\rho+3p\geq 0$, that in classical GR prevents any form of ``bouncing'' cosmology, as it is evident from one of the Friedman equations
\bea
{\ddot a\over a}=-{4\pi G\over 3}(\rho+3p).
\eea
In this equation, the sign of the cosmic acceleration $\ddot a$ is always negative because of the  energy condition. If this does not hold, the Universe can reverse the contraction into an expansion. Unfortunately, the violation of the energy condition occurs precisely at the energy scales at which the one-loop approximation, implicitly assumed in the semiclassical analysis, breaks down. These scales correspond roughly to the Planck energy density, where Eq.\ (\ref{semiclassical}) can no longer be trusted. Despite these drawbacks, semiclassical theory unveiled very important phenomena, such as Hawking radiation \cite{hawking1,hawking2} and the Unruh's effect \cite{unruh}. 

\subsection{Particle creation}

In cosmology, semiclassical theory is at the origin of the spontaneous particle creation in time-dependent backgrounds, discovered by Parker a long time ago \cite{parker1,parker2}. This effect is typical of quantum fields defined on time-dependent metrics, where there is no universal definition for a global time coordinate. In such a case, one needs to define the vacuum state of a field with respect to an asymptotically flat spacetime, say, in the infinite past. A simple calculation shows that the particle number is not invariant, and the vacuum state $|0\rangle_{\rm out}$ defined with respect to a future asymptotically flat spacetime does not coincide with the one defined in the past $|0\rangle_{\rm in}$, according to the scheme
\bea
|0\rangle_{\rm in}=\alpha|0\rangle_{\rm out}+\beta|0\rangle_{\rm out}.
\eea
In this expression, $\alpha$ and $\beta$ are called Bogolubov coefficients, and the particle number, measured by the ``in'' observer, is proportional to $|\beta|^{2}$. The evolution of the metric between the two asymptotically flat spacetime is assumed to be adiabatic, which, in cosmological context, means that the rate of expansion must be always smaller than the frequency of the modes excited. 
The most spectacular prediction derived from this phenomenon is the creation of scalar and tensor perturbations in the inflationary  Universe \cite{Mukhanov:1990me}, which are the seeds of the large scale structures seen today, and which are studied with increasing precision via the detection of  anisotropies in the Cosmic Microwave Background \cite{durrer}. In this respect, the observation of such fluctuations can be called  the first evidence for quantum gravitational effects \footnote{One should distinguish between relic gravitational waves, originated by quantum fluctuations of the metric, and gravitational waves emitted, for instance, by coalescing black holes. In the former case, we have a genuine quantum gravitational effect, while in the second the emission can be fully described by GR.}.

\subsection{Effects of the quantum origin of fluctuations}

There is a subtle question regarding the quantum origin of inflationary perturbations, which is usually overlooked.  Their spectra are conventionally constructed from the 2-point correlation function of a random variable $X(\vec k,t)$, which is regarded as a \emph{classical field}. This is  related to the power spectrum $P_{X}$ via the relation
\bea\label{power}
\langle X(\vec k,t)X(\vec k',t)\rangle=(2\pi)^{3}\delta(\vec k-\vec k')P_{X}(k)\ ,
\eea
where $k=|\vec k|$.

When we look at $X(\vec k,t)$ in terms of a \emph{quantum field} in momentum space, we need to reinterpret the average $\langle\ldots \rangle$ as the expectation value of the 2-point function over a determined quantum state, in the same way as in Eq.\ (\ref{semiclassical}). This apparently innocuous detail raises in fact several questions, usually ignored in the literature. For instance, the value of the expectation value depends upon the algebra of the annihilation and creation operators that form the field operator associated to $X(\vec k,t)$. Non-trivial algebra can lead to non trivial power spectra. Also, the quantum expectation value depends on the state of the field, and different choices can lead to radically different results.

Apart from these general aspects, the main problem with the quantum  origin of an expression like (\ref{power}),  is that it diverges for $\vec k\rightarrow \vec k'$. In classical analysis this problem is usually eliminated by introducing a UV cut-off by hand. But from the quantum field theory point of view, a simple UV cut-off would break Lorentz invariance. Fortunately, there are more sophisticated ways to deal with such UV divergences on curved space, that allows for a covariant regularization, that might have consequences on the observable spectrum. Suppose that $\varphi(\vec x,t)$ represents a perturbation propagating on an inflationary background. Upon quantization, we have
\bea
\hat\varphi(\vec x,t)=(2\pi)^{-3/2}\int d^{3}k\left[\varphi_{k}(t)\hat a_{\vec k}\,e^{i\vec k\cdot t}+\varphi_{k}^{*}(t)\hat a_{\vec k}^{\dagger}\,e^{-i\vec k\cdot t}\right]\ ,
\eea
where $\hat a_{\vec k}$ is the usual annihilation operator. The power spectrum $P^{2}$ is related to the 2-point function as the coincident limit
\bea
\langle \varphi^{2}(\vec x, t)\rangle=\int_{0}^{\infty} {dk\over k}P^{2}(k,t)\ .
\eea
When evaluated on a Robertson-Walker background of the form $ds^{2}=-dt^{2}+a^{2}(t)\delta_{ij}dx^{i}dx^{j}$, this quantity shows logarithmic and quadratic divergences as, in the large-$k$ limit one has
\bea
{P^{2}\over k}\sim{1\over 4\pi^{2}}\left({k\over a^2}+{\dot a^{2}\over a^{2}k}+\ldots\right)\ ,
\eea
independently of the exact form of the scale factor $a(t)$. While the first term corresponds to the usual flat-space divergence, the second one  arises because of the curvature of the background, and cannot be regularized by, for instance, normal ordering. L.\ Parker and collaborators proposed a way to cope with these divergences based on a well-known procedure, called adiabatic subtraction \cite{ParkerToms,BirrellandDavies}. Essentially, the idea is to subtract form the ``bare'' spectrum a counter-term $C_{k}$ determined by a WKB solution to the Klein-Gordon equation satisfied by the field $\varphi$. As a result, both ``renormalized''   scalar and tensor power spectra are UV finite and inevitably modified by the counter-terms themselves \cite{Agullo:2011qg,Agullo:2009zi,Agullo:2009vq,Parker:2007ni}.  For instance, the simplest chaotic inflationary models, with $\phi^{2}$ and $\phi^{4}$ potentials, are normally excluded by the analysis of the WMAP data. However, if one takes in account the adiabatic renormalization, it turns out that the spectra of these models are modified by the counter-term $C_{k}$ in such a way that it becomes  compatible with observations. These observational effects where questioned, however, by arguing that it is not clear whether adiabatic subtraction maintains its validity up the horizon exit of the relevant scales \cite{Marozzi:2011da,Durrer:2009ii}. Given that the adiabatic subtraction can alter significantly the interpretation of the observed spectra, it is worth keeping investigating on this topic. In particular, it would be crucial to test these results  by means of a different regularization method. In addition, It would be interesting  to study eventual non-gaussian imprints on the spectra.

\section{Towards the Planck scale}

As mentioned in the previous section, the semiclassical picture breaks down at the Planck scale. This leads to a potential problem as, in most inflationary models, quantum fluctuations responsible for scalar and tensor perturbations originate when the radius of the Universe is of the order of the Planck length $\ell_{p}=1.6\cdot10^{-35}$ m or even below. This is the so-called trans-Planckian problem in cosmology, a topic that has generated a flurry activity in the past decade.  The scalar and the tensor perturbation spectra are known to be almost flat, small deviations being caused by the weak time-dependence of the Hubble parameter during inflation. The naif expectation is that the unknown physics around the Planck scale leaves an imprint on the initial conditions for quantum fluctuations, and, therefore, on the spectra. There are several proposals that test this prediction, with some contrasting results that reveals that many aspects are still unclear.

\subsection{Modified dispersion relations}

In order to test the robustness of the flatness against trans-planckian modifications, some people modified the dispersion relation for modes with planckian energy, a way that has been inspired by similar investigations  in the context of ``dumb'' holes in condensed matter systems. The most important examples are the (generalized) Corley-Jacobson's dispersion relation \cite{Corley}
\bea
\omega^{2}(k)=k^{2}+\sum_{j=2}^{N}\alpha_{j}{k^{2j}},
\eea
and the Unruh's dispersion relation \cite{unruhdisp}
\bea
\omega(k)=k_{c}\tanh^{1/p} \left(k^{p}\over k_{c}^{p}\right)
\eea
where $\alpha_{j}$ and $k_{c}$ are parameters related to the scale at which Lorentz invariance breaks down, while $p$ is arbitrary. These dispersion relations break local Lorentz invariance, taking on board the possibility that the latter is no longer a fundamental symmetry of the underlying quantum gravity theory. Some authors  investigated the evolution of cosmological perturbations with modified dispersion via mode analysis \cite{Martin,Brandenberger}, other by computing the modified stress tensor and the associated backreaction \cite{Lemoine}. In both cases, a general agreement was found on the fact that distinct signatures of trans-planckian physics could appear in the spectra for certain kinds of dispersion relation and initial conditions. This conclusion was contrasted by arguing that relevant transplanckian effects would spoil the inflationary scenario itself, and so they must be excluded \cite{Staro}. The question is not settled yet, and more recently the attention has focussed on possible non-gaussian signatures of trans-planckian physics \cite{Chialva1,Chialva2,Ashoorioon:2011eg}. Another aspect of modified dispersion relation in semiclassical gravity is the possibility that the back-reaction can be larger that the usual case. This quite intricated issue has been investigated both in relation to inflationary cosmology \cite{Lopez Nacir:2007jx,Lopez Nacir:2007du} and to black holes \cite{Rinaldi:2008ep,Rinaldi:2007de}.

\subsection{Path integral duality}

So far, we discussed quantum gravity effects that violate local Lorentz invariance, but this is not the only possibility. In fact, one can introduce a modification of the field propagator that preserves Lorentz symmetry but eliminate the UV divergence at coincident points \cite{Agullo:2008qb}. This can be achieved in several ways. For instance, one can assume that the path integral amplitude is endowed with a ``duality'' symmetry \cite{Padmanabhan:1996ap,Padmanabhan:1998yya,Srinivasan:1997rs} that  maps any length $L<\ell_{p}$ to $L'=\ell_{p}/L$.  This hypothesis is motivated by the idea that spacetime has a discrete nature at very short distance, of the order of $\ell_{p}$, and it is somehow similar to the T-duality that maps type IIA to type IIB string theory. Concretely, path integral duality leads to a modification of the usual Feynman path integral for, say, a scalar field of mass $m$, according to 
\bea
G(x,y)=\sum \exp \left(-m\sigma(x,y)-{m\ell_{p}\over \sigma(x,y)}\right),
\eea
where $\sigma(x,y)$ is the Synge world function. As a result, the large momentum limit of the propagator reads
\bea
\tilde G(p)\sim {e^{-\ell_{p}\sqrt{p^{2}+m^{2}}}\over \ell^{1/2}(p^{2}+m^{2})^{3/4}},
\eea
which shows a suppression at transplanckian energies. The evolution of perturbations in the slow-roll inflationary scenario has been investigated and it reveals that the spectra remain scale-invariant, only their (unobservable) amplitude being modified \cite{Sriramkumar:2006qt}. 

\subsection{Non-commutative geometry}

In the Planckian phase of inflation, physics could be different in several ways, and modified dispersion relations are just one possibility to be explored. As mentioned in the previous section,
 the two-point function of a scalar field is constructed from basic quantum field theory, according to a set of rules determined in the context of relativistic quantum mechanics. In particular, the usual commutation rules between position and momentum are promoted to commutation rules between the field and its canonical conjugate. A modification of the fundamental quantum mechanical commutation rules can be easily generalized to field theory. The most popular case is represented by non-commutativity geometry, which implies that coordinate operators do not commute, i.e.
\bea
[\hat x^{\mu},\hat x^{\nu}]=i\theta^{\mu\nu}\ ,
\eea 
where $\theta^{\mu\nu}$ is an anti-symmetric matrix, usually taken to be constant \cite{ncfund}. There are many fundamental theories that phenomenologically reduce to an ordinary field theory over a non-commutative manifold, from string theory to loop quantum gravity. It is therefore important to consider the possibility that non-commutative effects  took place during the inflationary era, and to extract some observational signature.

At a fundamental level, one can construct a model where the inflationary expansion of the Universe  is driven by non-commutative effects on the matter fields \cite{mag}. In this kind of models, there is no need for an inflaton field, as  non-commutativity modifies the equation of state in the radiation-dominated Universe in a way that it generates a quasi-exponential expansion. The initial conditions are thermal and not determined by a quantum vacuum. The predictions for the power spectra have been worked out  by  Brandenberger and Koh \cite{koh}, who find that  the spectrum of fluctuations is nearly scale invariant, and shows a small red tilt, the magnitude of which is different from what is obtained in a usual inflationary model with the same expansion rate. 

\subsection{Coherent-state approach to inflation}

\begin{figure}
\centering
\includegraphics[width=8cm]{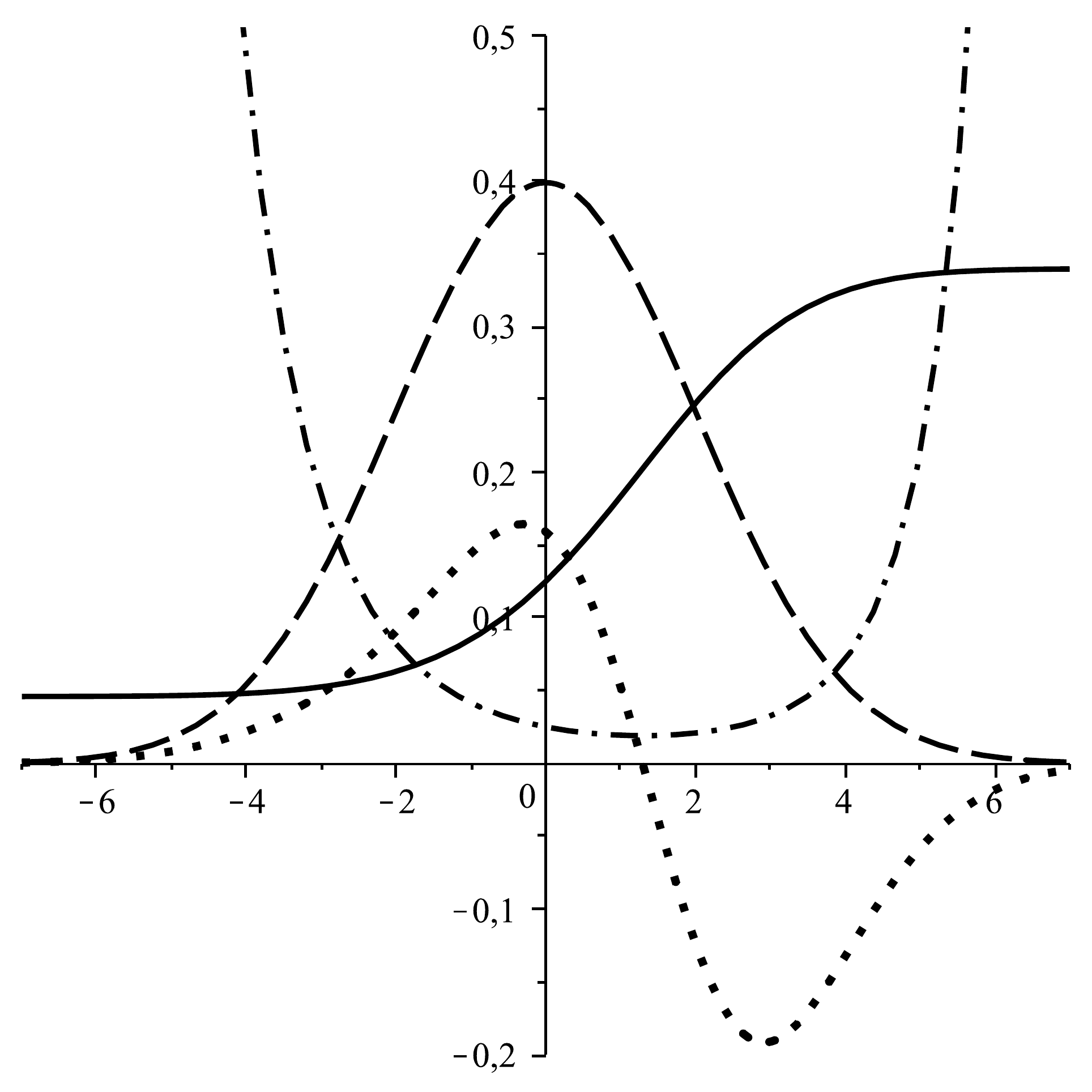} 
        \caption{Qualitative behaviour of $a$ (solid line), $\ddot a$ (dotted line), $H$ (dashed line), and $(aH)^{-1}$ (dot-dashed line) as functions of  time.}
	\label{FigA}
\end{figure}

An alternative point of view consists in following the semiclassical approach, by solving the cosmological equation of motions deriving from Eq.\ (\ref{semiclassical}), where now the expectation value is computed on a non-commutative background \cite{ncmax}. The advantage of this approach is that, thanks to non-commutativity, the expectation value is not divergent, hence there is no need to add complicated counter-terms. A realization of this idea is possible if based on the so-called coherent state non-commutativity \cite{Smailagic:2003rp,Smailagic:2003yb}. According to this construction, all coordinate  are promoted to operators that satisfy the relation $
[\hat z^{\mu},\hat z^{\nu}]=i\Theta^{\mu\nu} $, where $\Theta^{\mu\nu}$ is a constant and antisymmetric tensor. In  four Euclidean dimensions, one can transform this tensor in a block diagonal form, such that $\Theta^{\mu\nu}={\rm diag} (\theta_1 \varepsilon^{ij}, \theta_2 \varepsilon^{ij})$, where $\varepsilon^{ij}$ is the two-dimensional Levi-Civita tensor. It turns out that, if $\theta_1=\theta_2$, the resulting field theory is covariant \cite{Smailagic:2003rp,Smailagic:2003yb}. A further requirement is that physical coordinates are commuting c-numbers, constructed as expectation values on coherent states of $\hat z$. For example, on the Euclidean plane we have two coordinate operators, which satisfy the algebra $[\hat z_1,\hat z_2]=i\theta$. Then, one can construct the ladder operators
\bea
 \hat A=\hat z_1+i\hat z_2\ , \quad  \hat A^{\dagger}=\hat z_1-i\hat z_2\ ,
\eea
such that $[\hat A,\hat A^{\dagger}]=2\theta$. The coherent states $|\alpha\rangle$ are defined as the ones which satisfy the equation $\hat A |\alpha\rangle =\alpha |\alpha\rangle$. The physical coordinates are the \emph{commuting} $c$-numbers 
 \bea 
 x_1={\rm Re}(\alpha)=\langle \alpha | \hat z_1 |\alpha \rangle\ , \quad  x_2={\rm Im}(\alpha)=\langle \alpha | \hat z_2 |\alpha \rangle\ .
\eea
Thus, on the non-commutative plane, the vector $(x_1,x_2)$ represents the mean position of a point-particle. The above construction can be lifted to four-dimensional spacetimes, and it can be shown that the Euclidean propagator for a scalar field becomes
\bea
G_E(p)={e^{-p^2\theta/4}\over p^2+m^2}\ ,
\eea
displaying a Gaussian damping at high momenta. In coordinate space, the above propagator is UV finite, as can be seen from its expression in the massless limit \cite{Nicolini:2009dr,Rinaldi:2010zu}
\bea\label{propag}
G_E(x,x')={1-e^{-\sigma_E(x,x')/4\theta}\over 4\pi^2\sigma_E(x,x')}\ ,
\eea
where $\sigma_E$ is the Euclidean geodesic distance. As a consequence, also the stress tensor is UV finite. If matter in the early Universe is modeled by the usual perfect fluid, it turns out that the effective energy density reads
\bea
\langle\hat\rho\rangle= \rho_0\,e^{-(t-t_0)^2/4\theta}\ ,
\eea
an expression similar to the one found in a static and spherically symmetric background \cite{nico}. This leads to the non-singular Friedman equation (with $t_{0}=0$)
\bea\label{NCFried}
\left(\dot a\over a\right)^2={8\pi G\over 3}\rho(t)\equiv H_0^2\, e^{-t^2/4\theta}\ ,
\eea
where $H=\dot{a}/a$ and $a$ is the scale factor. As the energy density is no longer singular, we can extend $t$ from $+\infty$ to $-\infty$, and the scale factor reads
\bea\label{scaleF}
a(t)=a_0\exp\left[{H_0\sqrt{2\pi\theta}}\,\, {\rm Erf}\left(t\over 2\sqrt{2\theta}\right)\right]\ ,
\eea
where the error function is defined as
\bea
 {\rm Erf}(x')={2\over\sqrt{\pi}}\int_0^{x'}\,e^{-x^2}dx\ ,
\eea
and $a_0$ is an integration constant. The acceleration $\ddot a$ is initially positive and then changes sign at the time when the comoving Hubble length $(aH)^{-1}$ reaches its minimum, while $H$ reaches a maximum value at $t=0$, see Fig.\ (\ref{FigA}).  In other words, the global evolution of the scale factor shows a bounce, which is characterized only by the parameter $\theta$. In addition,  it can be shown  that $\sqrt{\theta} \sim 10^{-2}\cdot \ell_{p}$ if one asks for 60 e-folds of inflation  \cite{ncmax}. Although we would have expected a value for $\theta$ close to $\ell_{p}$, we find this result very encouraging and worth of further investigations. 
This model of non-commutative inflationary Universe is quite intriguing also on another aspect. The leading order for the energy density is a constant, which, if interpreted as a cosmological constant, it gives a way too large value if $\theta$ is of the order of magnitude found above. The situation can be improved if one admits that also momenta do not commute, according to the generalized operator algebra \cite{NCmom1,NCmom2}
 \bea
[\hat x_i,\hat x_j]=i\theta\varepsilon_{ij} \ ,\quad [\hat p_i,\hat p_j]=i\gamma\varepsilon_{ij} \ , \quad [\hat x_i,\hat p_j]=i\hbar\delta_{ij}\ ,
\eea
where $\varepsilon_{ij} $ is the two-dimensional Levi-Civita tensor, and $\gamma$ is a sort of ``minimal momentum''. This algebra can be reduced to the one with commuting momenta by rescaling the latter according to
\bea
\hat p_i\longrightarrow \hat \pi_i=\alpha \hat p_i+\beta \varepsilon_{ij} \hat x_j\ , \quad \alpha={1\over \sqrt{1-\theta\gamma}}\ , \\\non
 \beta={1\over\theta}(1-\alpha) \ ,\quad [\hat\pi_i,\hat\pi_j]=0\ .
\eea
Note that, at the value $\gamma\theta=1$,  the linear transformation above becomes singular. This situation corresponds to a change of the symmetry group acting on the plane, from SU$(2)$ to SU$(1,1)$ \cite{NCmom2}. The important fact for the inflationary scenario, is the expectation value on coherent states of the rescaled wave operator $\exp(i\hat{\vec{\pi}}\cdot \hat {\vec x})$ yields a damping factor multiplied by $\alpha$, and the energy momentum tensor becomes
\bea
\langle T_{\mu\nu}\rangle\sim{1\over\theta^2\alpha^4}g_{\mu\nu}={(1-\gamma\theta)^2\over \theta^2}g_{\mu\nu}\ .
\eea
This shows that the effective cosmological constant can actually be much smaller than previously estimated, if the product $\gamma\theta$ is close to one. As the true cosmological constant is very small, it is intriguing to argue that it is related to some kind of phase transition, similar to the one linked to the   SU$(2)$ to SU$(1,1)$ symmetry change cited above.

\subsection{Observational signatures of non-commutativity}

So far, there has been little work on the experimental detection of non-commutative effects in the inflationary Universe. Recently, it was found that, in the case where scalar and tensor perturbations generated during inflation  are subjected to non-commutativity at short distance,  the power spectrum is modified according to  \cite{Koivi}
\bea
P=P_{0}\,e^{H\vec\theta\cdot \vec k}\ ,
\eea
where $H$ is the Hubble parameter, $P_{0}$ is the usual commutative spectrum, and $\vec \theta$ is the vector formed by the $\theta^{0i}$ components of $\theta^{\mu\nu}$. The most important aspect is therefore that the spectrum becomes direction-dependent. Moreover,  also specific non-gaussian signatures can arise \cite{Koivi}. If the direction dependence of the spectrum will be measured by future data, it will be possible to determine the non-commutative scale $\theta$. In particular, the dipole modulation seems to be compatible with the recently observed hemispherical power asymmetry \cite{Koivi2}.

\section{Conclusion}

In this review we discussed several ideas dealing with the problem of the consistent quantization of gravitational interactions in the context of cosmology. In particular, we focussed on models based on the canonical quantization of fields on curved background, and on the most recent developments of this method. This is probably the most conservative approach, but, at the moment, it is also better testable than other more radical ideas. 

We are entering the era of precision cosmology, with the possibility of measuring cosmological parameters with an unthinkable accuracy just up to few years ago. Thus, the need for theoretical support to correctly interpret the data becomes more and more stringent. Current and future observations might offer the first glimpse of quantum gravitational effects, provided these are clearly disentangled among them and from other effects. 

Recent developments seem to indicate that string cosmology observational predictions are in contrast with observations. However, this is not the case of LQG as, so far, all calculations have shown that predictions are consistent with data. These calculations are promising and encouraging, and, in fact, a growing attention is being devoted to LQG phenomenology by main-stream cosmologists.

Generally speaking, the priority is to work out precisely the spectra and bi-spectra associated to tensor and scalar fluctuations emerging from the quantum models of the early Universe discussed in this review. Only in this way we will not miss the opportunity to see, for the first time, a glimpse of the quantum origin of our Universe.

%\section*{Acknowledgments}

%%%%%%%%%%%%%%


\begin{thebibliography}{0}

%%%% hawk ellis %%%

\bibitem{Hawking:1973uf}
  S.~W.~Hawking and G.~F.~R.~Ellis,
  ``The Large scale structure of space-time,''
  Cambridge University Press, Cambridge, 1973
%%%% mukhanov %%%

\bibitem{Mukhanov:2005sc}
  V.~Mukhanov,
  ``Physical foundations of cosmology,''
  Cambridge, UK: Univ. Pr. (2005) 421 p
  
%%%%%%%%% euclid and planck %%%%%

  \bibitem{sitePL}
  http://www.rssd.esa.int/index.php?project=Planck

\bibitem{EditorialTeam:2011mu}
  Editorial Team, R.~Laureijs, J.~Amiaux, S.~Arduini, J.~-L.~Augueres, J.~Brinchmann, R.~Cole and M.~Cropper {\it et al.},
  ``Euclid Definition Study Report,''
  arXiv:1110.3193 [astro-ph.CO].
  %%CITATION = ARXIV:1110.3193;%%
 
\bibitem{site}
http://www.euclid-ec.org/

%%%%%%%%% string cosmology %%%%%%%%%%%

\bibitem{Gasperini:2002bn}
  M.~Gasperini and G.~Veneziano,
  %``The pre-big bang scenario in string cosmology,''
  Phys.\ Rept.\  {\bf 373} (2003) 1
  [arXiv:hep-th/0207130].
  %%CITATION = PRPLC,373,1;%%
  
  
  
 
 
   
  
 %%%%%%% RS model and generalizations %%%%%%% 
 
 %%review
 
 \bibitem{Brax:2004xh}
  P.~Brax, C.~van de Bruck and A.~-C.~Davis,
  %``Brane world cosmology,''
  Rept.\ Prog.\ Phys.\  {\bf 67} (2004) 2183
  [hep-th/0404011].
  %%CITATION = HEP-TH/0404011;%%
  
 %%review
  
 
 
  \bibitem{Randall:1999ee}
  L.~Randall and R.~Sundrum,
  %``A Large mass hierarchy from a small extra dimension,''
  Phys.\ Rev.\ Lett.\  {\bf 83} (1999) 3370
  [hep-ph/9905221].
  %%CITATION = HEP-PH/9905221;%%
  \bibitem{Randall:1999vf}
  L.~Randall and R.~Sundrum,
  %``An Alternative to compactification,''
  Phys.\ Rev.\ Lett.\  {\bf 83} (1999) 4690
  [hep-th/9906064].
  %%CITATION = HEP-TH/9906064;%%
  
   \bibitem{Birmingham:2001dq}
  D.~Birmingham and M.~Rinaldi,
  %``Brane world in a topological black hole bulk,''
  Mod.\ Phys.\ Lett.\ A {\bf 16} (2001) 1887
  [hep-th/0106237].
  %%CITATION = HEP-TH/0106237;%%
  \bibitem{Kraus:1999it}
  P.~Kraus,
  %``Dynamics of anti-de Sitter domain walls,''
  JHEP {\bf 9912} (1999) 011
  [hep-th/9910149].
  %%CITATION = HEP-TH/9910149;%%
  
  % implementation with T duality
  
  \bibitem{Corradini:2005pm}
  O.~Corradini and M.~Rinaldi,
  %``Self-T-dual brane cosmology and the cosmological constant problem,''
  JCAP {\bf 0601} (2006) 020
  [hep-th/0509200].
  %%CITATION = HEP-TH/0509200;%%
 \bibitem{Rinaldi:2004hh}
  M.~Rinaldi and P.~Watts,
  %``Pre-big bang scenario on self-T-dual bouncing branes,''
  JCAP {\bf 0503} (2005) 006
  [hep-th/0411185].
  %%CITATION = HEP-TH/0411185;%%
 \bibitem{Rinaldi:2003is}
  M.~Rinaldi,
  %``Brane worlds in T dual bulks,''
  Phys.\ Lett.\ B {\bf 582} (2004) 249
  [hep-th/0311147].
 %%CITATION = HEP-TH/0311147;%%
  
  


 
  
%%%%%%%%%%% loop quantum gravity %%%%%%%

\bibitem{dewitt}
B.\ S.\ DeWitt, Phys.\ Rev.\ {\bf 160} 1113Ð1148 (1967)
\bibitem{wheeler}
J.\ A.\ Wheeler, ``Superspace and quantum geometrodynamics'', in Battelle Rencontres, edited by J. A.\ Wheeler and C.\ M.\ DeWitt (W.\ A.\ Benjamin, New York, 1972).
\bibitem{Banerjee:2011qu}
  K.~Banerjee, G.~Calcagni and M.~Martin-Benito,
  %``Introduction to loop quantum cosmology,''
  arXiv:1109.6801 [gr-qc].
  %%CITATION = ARXIV:1109.6801;%%
\bibitem{Ashtekar:2011ni}
  A.~Ashtekar and P.~Singh,
  %``Loop Quantum Cosmology: A Status Report,''
  Class.\ Quant.\ Grav.\  {\bf 28} (2011) 213001
  [arXiv:1108.0893 [gr-qc]].
  %%CITATION = ARXIV:1108.0893;%%
\bibitem{Bojowald:2011zz}
  M.~Bojowald, (ed.),
  %``Quantum cosmology,''
  Lect.\ Notes Phys.\  {\bf 835} (2011) 1.
  %%CITATION = LNPHA,835,1;%%
  
  
%%%%%%%%% horava lifschitz  %%%%%%%%%%
  
%% original

\bibitem{Horava:2009uw}
  P.~Horava,
  %``Quantum Gravity at a Lifshitz Point,''
  Phys.\ Rev.\ D {\bf 79} (2009) 084008
  [arXiv:0901.3775 [hep-th]].
  %%CITATION = ARXIV:0901.3775;%%
  
%%%% inconsistencies

\bibitem{Koyama:2009hc}
  K.~Koyama and F.~Arroja,
  %``Pathological behaviour of the scalar graviton in Horava-Lifshitz gravity,''
  JHEP {\bf 1003} (2010) 061
  [arXiv:0910.1998 [hep-th]].
  %%CITATION = ARXIV:0910.1998;%%
  \bibitem{Blas:2009yd}
  D.~Blas, O.~Pujolas and S.~Sibiryakov,
  %``On the Extra Mode and Inconsistency of Horava Gravity,''
  JHEP {\bf 0910} (2009) 029
  [arXiv:0906.3046 [hep-th]].
  %%CITATION = ARXIV:0906.3046;%%


%% extended

\bibitem{Blas:2009qj}
  D.~Blas, O.~Pujolas and S.~Sibiryakov,
  %``Consistent Extension of Horava Gravity,''
  Phys.\ Rev.\ Lett.\  {\bf 104} (2010) 181302
  [arXiv:0909.3525 [hep-th]].
  %%CITATION = ARXIV:0909.3525;%%
  
  
%%%% aether %%%%
\bibitem{Jacobson:2010mx}
  T.~Jacobson,
  %``Extended Horava gravity and Einstein-aether theory,''
  Phys.\ Rev.\ D {\bf 81} (2010) 101502
   [Erratum-ibid.\ D {\bf 82} (2010) 129901]
  [arXiv:1001.4823 [hep-th]].
  %%CITATION = ARXIV:1001.4823;%%

  
%%%%% HL cosmology %%%%%%%

\bibitem{Calcagni:2009ar}
  G.~Calcagni,
  %``Cosmology of the Lifshitz universe,''
  JHEP {\bf 0909} (2009) 112
  [arXiv:0904.0829 [hep-th]].
  %%CITATION = ARXIV:0904.0829;%%
  \bibitem{Brandenberger:2009yt}
  R.~Brandenberger,
  %``Matter Bounce in Horava-Lifshitz Cosmology,''
  Phys.\ Rev.\ D {\bf 80} (2009) 043516
  [arXiv:0904.2835 [hep-th]].
  %%CITATION = ARXIV:0904.2835;%%

 \bibitem{Mukohyama:2009gg}
  S.~Mukohyama,
  %``Scale-invariant cosmological perturbations from Horava-Lifshitz gravity without inflation,''
  JCAP {\bf 0906} (2009) 001
  [arXiv:0904.2190 [hep-th]].
  %%CITATION = ARXIV:0904.2190;%% 

%%%%%%%%%% two to four dimensional transition %%%%%%%%%

\bibitem{Horava:2009if}
  P.~Horava,
  %``Spectral Dimension of the Universe in Quantum Gravity at a Lifshitz Point,''
  Phys.\ Rev.\ Lett.\  {\bf 102} (2009) 161301
  [arXiv:0902.3657 [hep-th]].
  %%CITATION = ARXIV:0902.3657;%%

\bibitem{Carlip:2009km}
  S.~Carlip,
  %``The Small Scale Structure of Spacetime,''
  arXiv:1009.1136 [gr-qc].
  %%CITATION = ARXIV:1009.1136;%%  
\bibitem{Carlip:2009kf}
  S.~Carlip,
  %``Spontaneous Dimensional Reduction in Short-Distance Quantum Gravity?,''
  arXiv:0909.3329 [gr-qc].
  %%CITATION = ARXIV:0909.3329;%%  
  \bibitem{Carlip:2011tt}
  S.~Carlip, R.~A.~Mosna and J.~P.~M.~Pitelli,
  %``Vacuum Fluctuations and the Small Scale Structure of Spacetime,''
  Phys.\ Rev.\ Lett.\  {\bf 107} (2011) 021303
  [arXiv:1103.5993 [gr-qc]].
  %%CITATION = ARXIV:1103.5993;%%


  
  
 %%%%%%%%% observational signatures of two to four dim transition %%%%%%%%%%%
 
 \bibitem{Mureika:2011bv}
  J.~R.~Mureika and D.~Stojkovic,
  %``Detecting Vanishing Dimensions Via Primordial Gravitational Wave Astronomy,''
  Phys.\ Rev.\ Lett.\  {\bf 106} (2011) 101101
  [arXiv:1102.3434 [gr-qc]].
  %%CITATION = ARXIV:1102.3434;%%
\bibitem{Rinaldi:2010yp}
  M.~Rinaldi,
  %``Observational signatures of pre-inflationary and lower-dimensional effective gravity,''
  arXiv:1011.0668 [astro-ph.CO].
  %%CITATION = ARXIV:1011.0668;%%
  

%%%%%%%%%%% space-time thermodynamics %%%%%%

%%%%%%%%%%% space-time thermodynamics %%%%%%

\bibitem{Jacobson:1995ab}
  T.~Jacobson,
  %``Thermodynamics of space-time: The Einstein equation of state,''
  Phys.\ Rev.\ Lett.\  {\bf 75} (1995) 1260
  [gr-qc/9504004].
  %%CITATION = GR-QC/9504004;%%

  \bibitem{Padmanabhan:2003gd}
  T.~Padmanabhan,
  %``Gravity and the thermodynamics of horizons,''
  Phys.\ Rept.\  {\bf 406} (2005) 49
  [gr-qc/0311036].
  %%CITATION = GR-QC/0311036;%%
 



  

  
   
  %%%%%%%%%%%%%%%%%% QFT on curved space  books %%%%%%%%%%
  
 \bibitem{ParkerToms}
L.\ Parker  and D.\ J.\ Toms, ``Quantum field theory in curved
spacetime: quantized Þelds and gravity'', CUP (2009)

\bibitem{BirrellandDavies}
N.\ D.\ Birrell and P.\ C.\ W.\ Davies, ``Quantum fields in curved
space'', Cambridge University Press, (1982).

 

\bibitem{hawking1} S.\ W.\ Hawking, Nature (London) {\bf 248}, 30 (1974).
  
\bibitem{hawking2} S.\ W.\ Hawking, Commun.\ Math.\ Phys.\ {\bf 43}, 199 (1975).



\bibitem{unruh}
  W.~G.~Unruh,
  %``Notes on black hole evaporation,''
  Phys.\ Rev.\ D {\bf 14} (1976) 870.
  %%CITATION = PHRVA,D14,870;%%
  
  %%%%%%%%%%%%%%%%%% primordial fluctuations %%%%%%%%%%

\bibitem{parker1}
  L.~Parker,
  %``Quantized fields and particle creation in expanding universes. 1.,''
  Phys.\ Rev.\  {\bf 183} (1969) 1057.
  %%CITATION = PHRVA,183,1057;%%
\bibitem{parker2}
  L.~Parker,
  %``Quantized fields and particle creation in expanding universes. 2.,''
  Phys.\ Rev.\ D {\bf 3} (1971) 346
   [Erratum-ibid.\ D {\bf 3} (1971) 2546].
  %%CITATION = PHRVA,D3,346;%%  
  
\bibitem{Mukhanov:1990me}
  V.~F.~Mukhanov, H.~A.~Feldman and R.~H.~Brandenberger,
  %``Theory of cosmological perturbations. Part 1. Classical perturbations. Part 2. Quantum theory of perturbations. Part 3. Extensions,''
  Phys.\ Rept.\  {\bf 215} (1992) 203.
  %%CITATION = PRPLC,215,203;%%
  
\bibitem{durrer}
R.\ Durrer, ``The Cosmic Microwave Background'', Cambridge University Press, Cambridge, England, 2008.


%%%%%%%% normalization of cosm perturbations %%%%%%%


\bibitem{Agullo:2011qg}
  I.~Agullo, J.~Navarro-Salas, G.~J.~Olmo and L.~Parker,
  %``Remarks on the renormalization of primordial cosmological perturbations,''
  Phys.\ Rev.\ D {\bf 84} (2011) 107304
  [arXiv:1108.0949 [gr-qc]].
  %%CITATION = ARXIV:1108.0949;%%
  
  \bibitem{Agullo:2009zi}
  I.~Agullo, J.~Navarro-Salas, G.~J.~Olmo and L.~Parker,
  %``Revising the observable consequences of slow-roll inflation,''
  Phys.\ Rev.\ D {\bf 81} (2010) 043514
  [arXiv:0911.0961 [hep-th]].
  %%CITATION = ARXIV:0911.0961;%%
  
  \bibitem{Agullo:2009vq}
  I.~Agullo, J.~Navarro-Salas, G.~J.~Olmo and L.~Parker,
  %``Revising the predictions of inflation for the cosmic microwave background anisotropies,''
  Phys.\ Rev.\ Lett.\  {\bf 103} (2009) 061301
  [arXiv:0901.0439 [astro-ph.CO]].
  %%CITATION = ARXIV:0901.0439;%%
  
  \bibitem{Parker:2007ni}
  L.~Parker,
  %``Amplitude of Perturbations from Inflation,''
  hep-th/0702216 [HEP-TH].
  %%CITATION = HEP-TH/0702216;%%


  
    
  \bibitem{Marozzi:2011da}
  G.~Marozzi, M.~Rinaldi and R.~Durrer,
  %``On infrared and ultraviolet divergences of cosmological perturbations,''
  Phys.\ Rev.\ D {\bf 83} (2011) 105017
  [arXiv:1102.2206 [astro-ph.CO]].
  %%CITATION = ARXIV:1102.2206;%%
  
  \bibitem{Durrer:2009ii}
  R.~Durrer, G.~Marozzi and M.~Rinaldi,
  %``On Adiabatic Renormalization of Inflationary Perturbations,''
  Phys.\ Rev.\ D {\bf 80} (2009) 065024
  [arXiv:0906.4772 [astro-ph.CO]].
  %%CITATION = ARXIV:0906.4772;%%



  %%%%%%%%%%%%%%%%%% TP problem %%%%%%%%%%


\bibitem{Corley}
  S.~Corley and T.~Jacobson,
  %``Hawking spectrum and high frequency dispersion,''
  Phys.\ Rev.\ D {\bf 54} (1996) 1568
  [hep-th/9601073].
  %%CITATION = HEP-TH/9601073;%%
  
 \bibitem{unruhdisp}
W.~G.~Unruh,
  ``Dumb holes and the effects of high frequencies on black hole evaporation,''
  gr-qc/9409008.
  %%CITATION = GR-QC/9409008;%%
 
 


\bibitem{Martin}
  J.~Martin and R.~H.~Brandenberger,
  %``The TransPlanckian problem of inflationary cosmology,''
  Phys.\ Rev.\ D {\bf 63} (2001) 123501
  [hep-th/0005209].
  %%CITATION = HEP-TH/0005209;%%
\bibitem{Brandenberger}
R.~H.~Brandenberger and J.~Martin,
  %``The Robustness of inflation to changes in superPlanck scale physics,''
  Mod.\ Phys.\ Lett.\ A {\bf 16} (2001) 999
  [astro-ph/0005432].
  %%CITATION = ASTRO-PH/0005432;%%
  
  \bibitem{Lemoine}
  M.~Lemoine, M.~Lubo, J.~Martin and J.~-P.~Uzan,
  %``The Stress energy tensor for transPlanckian cosmology,''
  Phys.\ Rev.\ D {\bf 65} (2002) 023510
  [hep-th/0109128].
  %%CITATION = HEP-TH/0109128;%%
  
  \bibitem{Staro}
  A.~A.~Starobinsky,
  %``Robustness of the inflationary perturbation spectrum to transPlanckian physics,''
  Pisma Zh.\ Eksp.\ Teor.\ Fiz.\  {\bf 73} (2001) 415
   [JETP Lett.\  {\bf 73} (2001) 371]
  [astro-ph/0104043].
  %%CITATION = ASTRO-PH/0104043;%%
  


\bibitem{Chialva1}
  D.~Chialva,
  %``Signatures of very high energy physics in the squeezed limit of the bispectrum from the field theoretical approach,''
  arXiv:1108.4203 [astro-ph.CO].
  %%CITATION = ARXIV:1108.4203;%%
  
\bibitem{Chialva2}
  D.~Chialva,
  %``Enhanced CMBR non-Gaussianities from Lorentz violation,''
  JCAP {\bf 1201} (2012) 037
  [arXiv:1106.0040 [hep-th]].
  %%CITATION = ARXIV:1106.0040;%%  

\bibitem{Ashoorioon:2011eg}
  A.~Ashoorioon, D.~Chialva and U.~Danielsson,
  %``Effects of Nonlinear Dispersion Relations on Non-Gaussianities,''
  JCAP {\bf 1106} (2011) 034
  [arXiv:1104.2338 [hep-th]].
  %%CITATION = ARXIV:1104.2338;%%
  
  
  
  %%%%%%%%%%% modified dispersion relations in semiclassical approach %%%%%%%%%

\bibitem{Lopez Nacir:2007jx}
  D.~Lopez Nacir and F.~D.~Mazzitelli,
  %``Backreaction in trans-Planckian cosmology: Renormalization, trace anomaly and selfconsistent solutions,''
  Phys.\ Rev.\ D {\bf 76} (2007) 024013
  [arXiv:0706.2179 [gr-qc]].
  %%CITATION = ARXIV:0706.2179;%%
  \bibitem{Lopez Nacir:2007du}
  D.~Lopez Nacir, F.~D.~Mazzitelli and C.~Simeone,
  %``Adiabatic renormalization in theories with modified dispersion relations,''
  J.\ Phys.\ A A {\bf 40} (2007) 6895
  [gr-qc/0703016].
  %%CITATION = GR-QC/0703016;%%
\bibitem{Rinaldi:2008ep}
  M.~Rinaldi,
  %``A Momentum-space representation of Green's functions with modified dispersion relations on general backgrounds,''
  Phys.\ Rev.\ D {\bf 78} (2008) 024025
  [arXiv:0803.3684 [gr-qc]].
  %%CITATION = ARXIV:0803.3684;%%
\bibitem{Rinaldi:2007de}
  M.~Rinaldi,
  %``A Momentum-space representation of Green's functions with modified dispersion on ultra-static space-time,''
  Phys.\ Rev.\ D {\bf 76} (2007) 104027
  [arXiv:0709.2657 [gr-qc]].
  %%CITATION = ARXIV:0709.2657;%%

 

  
%%%%%%%%%% L invariant TP problem %%%%%%%%

\bibitem{Agullo:2008qb}
  I.~Agullo, J.~Navarro-Salas, G.~J.~Olmo and L.~Parker,
  %``Two-point functions with an invariant Planck scale and thermal effects,''
  Phys.\ Rev.\ D {\bf 77} (2008) 124032
  [arXiv:0804.0513 [hep-th]].
  %%CITATION = ARXIV:0804.0513;%%  



\bibitem{Padmanabhan:1996ap}
  T.~Padmanabhan,
  %``Duality and zero point length of space-time,''
  Phys.\ Rev.\ Lett.\  {\bf 78} (1997) 1854
  [hep-th/9608182].
  %%CITATION = HEP-TH/9608182;%%
  
\bibitem{Padmanabhan:1998yya}
  T.~Padmanabhan,
  %``Hypothesis of path integral duality. 1. Quantum gravitational corrections to the propagator,''
  Phys.\ Rev.\ D {\bf 57} (1998) 6206.
  %%CITATION = PHRVA,D57,6206;%%

\bibitem{Srinivasan:1997rs}
  K.~Srinivasan, L.~Sriramkumar and T.~Padmanabhan,
  %``The Hypothesis of path integral duality. 2. Corrections to quantum field theoretic results,''
  Phys.\ Rev.\ D {\bf 58} (1998) 044009
  [gr-qc/9710104].
  %%CITATION = GR-QC/9710104;%%


\bibitem{Sriramkumar:2006qt}
  L.~Sriramkumar and S.~Shankaranarayanan,
  %``Path integral duality and Planck scale corrections to the primordial
  %spectrum in exponential inflation,''
  JHEP {\bf 0612} (2006) 050
  [arXiv:hep-th/0608224].
  %%CITATION = JHEPA,0612,050;%%
  
  
  
 %%%%%%%% NC stuff %%%%%%%%%%
 
 \bibitem{ncfund}
H.\ S.\ Snyders, Phys.\ Rev.\ {\bf 71}, 38 (1947);
A. Connes, {\it G\'eometrie non commutative}, (InterEditions, Paris, France, 1990). 

\bibitem{mag}
S.~Alexander, R.~Brandenberger and J.~Magueijo,
  %``Non-commutative inflation,''
  Phys.\ Rev.\  D {\bf 67} (2003) 081301.
  %%CITATION = PHRVA,D67,081301;%%
  

  

\bibitem{koh}
S.~Koh and R.~H.~Brandenberger,
  %``Cosmological perturbations in non-commutative inflation,''
  JCAP {\bf 0706} (2007) 021.
  %%CITATION = JCAPA,0706,021;%%


 




  
\bibitem{ncmax}
  M.~Rinaldi,
  %``A New approach to non-commutative inflation,''
  Class.\ Quant.\ Grav.\  {\bf 28} (2011) 105022
  [arXiv:0908.1949 [gr-qc]].
  %%CITATION = ARXIV:0908.1949;%%
  
\bibitem{Smailagic:2003rp}
  A.~Smailagic and E.~Spallucci,
  %``UV divergence free QFT on noncommutative plane,''
  J.\ Phys.\ A A {\bf 36} (2003) L517
  [hep-th/0308193].
  %%CITATION = HEP-TH/0308193;%%  
  
\bibitem{Smailagic:2003yb}
  A.~Smailagic and E.~Spallucci,
  %``Feynman path integral on the noncommutative plane,''
  J.\ Phys.\ A A {\bf 36} (2003) L467
  [hep-th/0307217].
  %%CITATION = HEP-TH/0307217;%%  

    
  \bibitem{Rinaldi:2010zu}
  M.~Rinaldi,
  %``Particle Production and Transplanckian Problem on the Non-Commutative Plane,''
  Mod.\ Phys.\ Lett.\ A {\bf 25} (2010) 2805
  [arXiv:1003.2408 [hep-th]].
  %%CITATION = ARXIV:1003.2408;%%
  

\bibitem{nico}
  P.~Nicolini,
  Int.\ J.\ Mod.\ Phys.\ A {\bf 24}, 1229 (2009)
  [arXiv:0807.1939 [hep-th]].
  %%CITATION = ARXIV:0807.1939;%%
  
  \bibitem{Nicolini:2009dr}
  P.~Nicolini and M.~Rinaldi,
  %``A Minimal length versus the Unruh effect,''
  Phys.\ Lett.\ B {\bf 695} (2011) 303
  [arXiv:0910.2860 [hep-th]].
  %%CITATION = ARXIV:0910.2860;%%

  
 \bibitem{NCmom1}
V.~P.~Nair and A.~P.~Polychronakos,
  %``Quantum mechanics on the noncommutative plane and sphere,''
  Phys.\ Lett.\  B {\bf 505} (2001) 267;
  %%CITATION = PHLTA,B505,267;%%
  
  \bibitem{NCmom2}
  S.~Bellucci, A.~Nersessian and C.~Sochichiu,
  %``Two phases of the non-commutative quantum mechanics,''
  Phys.\ Lett.\  B {\bf 522} (2001) 345.
  %%CITATION = PHLTA,B522,345;%%
  
  \bibitem{Koivi}
  T.~S.~Koivisto and D.~F.~Mota,
  ``CMB statistics in noncommutative inflation,''
  arXiv:1011.2126 [astro-ph.CO].
  %%CITATION = ARXIV:1011.2126;%%

\bibitem{Koivi2}
N.~E.~Groeneboom, M.~Axelsson, D.~F.~Mota and T.~Koivisto,
  ``Imprints of a hemispherical power asymmetry in the seven-year WMAP data due
  to non-commutativity of space-time,''
  arXiv:1011.5353 [astro-ph.CO].
  %%CITATION = ARXIV:1011.5353;%%


\end{thebibliography}
\end{document}